# onepot CORE — an enumerated chemical space to streamline drug discovery, enabled by automated small molecule synthesis and AI

Andrei S. Tyrin,[1] Brandon Wang,[1] Manuel Muñoz,[1] Samuel H. Foxman,[1] Daniil A. Boiko[1,*]

[1] Onepot AI, Inc., 630 Gateway Blvd, South San Francisco, California, United States of America
*daniil@onepot.ai

## Abstract

The design–make–test–analyze cycle in early-stage drug discovery remains constrained primarily by the "make" step: small-molecule synthesis is slow, costly, and difficult to scale or automate across diverse chemotypes. Enumerated chemical spaces aim to reduce this bottleneck by predefining synthesizable regions of chemical space from available building blocks and reliable reactions, yet existing commercial spaces are still limited by long turnaround times, narrow reaction scope, and substantial manual decision-making in route selection and execution.

Here we present the first version of onepot CORE, an enumerated chemical space containing 3.4B molecules and corresponding on-demand synthesis product enabled by an automated synthesis platform and an AI chemist, *Phil*, that designs, executes, and analyzes experiments. onepot CORE is constructed by (i) selecting a reaction set commonly used in medicinal chemistry, (ii) sourcing and curating building blocks from supplier catalogs, (iii) enumerating candidate products, and (iv) applying ML-based feasibility assessment to prioritize compounds for robust execution. In the current release, the space is supported by seven reactions.

We describe an end-to-end workflow — from route selection and automated liquid handling through workup and purification. We further report validation across operational metrics (success rate, timelines, purity, and identity), including NMR confirmation for a representative set of synthesized compounds and assay suitability demonstrated using a series of DPP4 inhibitors. Collectively, onepot CORE illustrates a path toward faster, more reliable access to diverse small molecules, supporting accelerated discovery in pharmaceuticals and beyond.

## Introduction

Drug discovery is a complex, iterative process, that can be very broadly described as DMTA cycle — design, make, test, and analyze. Many research groups and companies made significant efforts in improving design step of the process. Testing, when done in-house, is usually very fast. Ways of analyzing the results are established as well. The entire cycle is mostly constrained by making compounds.[1] While synthesis can be easily automated for peptides, it is notoriously hard for small molecules.

Enumerated chemical spaces were designed to address this problem.[2] It is widely recognized that initial hits serve as chemical starting points rather than clinical candidates. So instead of trying to get "that particular structure", researchers take available reagents, a set of good working, reliable reactions, and enumerate all possible compounds that can be synthesized.[3] This idea became a foundation of successful products, such as Enamine REAL,[4] WuXi GalaXi, and the Freedom Space.[5]

However, arguably, these spaces still have a lot of room for growth (**Figure 1a**). Delivery times are on order of weeks,[3] often averaging at 4-5 weeks of synthesis times. These spaces also do not expand as much as they could have due to manual nature of the entire process, with no clear avenue for adding more reactions without significant human involvement in sight.

One of the reasons for this is still substantially manual character of the work. While research on automated chemical synthesis spans more than half a century,[6–8] little results of these works were adopted in industry. Furthermore, even if one could have an automated synthesis system, decision making would still be a bottleneck. This decision-making aspect of synthesis automation is largely ignored.

onepot recently launched with its AI chemist *Phil* — an AI-system capable of not only searching literature and analyzing results[9] but also designing and executing physical experiments (**Figure 1b**).[10] Furthermore, on analysis side, *Phil* can use analytical data to identify byproducts and develop strategies to mitigate their formation. Such a system would be indispensable for autonomous expansion of chemical spaces.

Here, we present onepot CORE (**C**ompounds **O**n-demand via **R**obotic **E**xecution) — an enumerated chemical space that underpins corresponding commercial product, providing chemical synthesis as fast as in five business days with delivery in the United States with median time just under ten business days. Here, we describe the strategy of the space construction, outline synthesis procedure to further illustrate the features of the final product, provide overview of the space, comparing it to competitive spaces, provide commercial testing statistics, and perform additional validation experiments. The team strongly believes that democratization of small molecule synthesis will support discoveries of novel pharmaceuticals, organic materials, and fragrances.

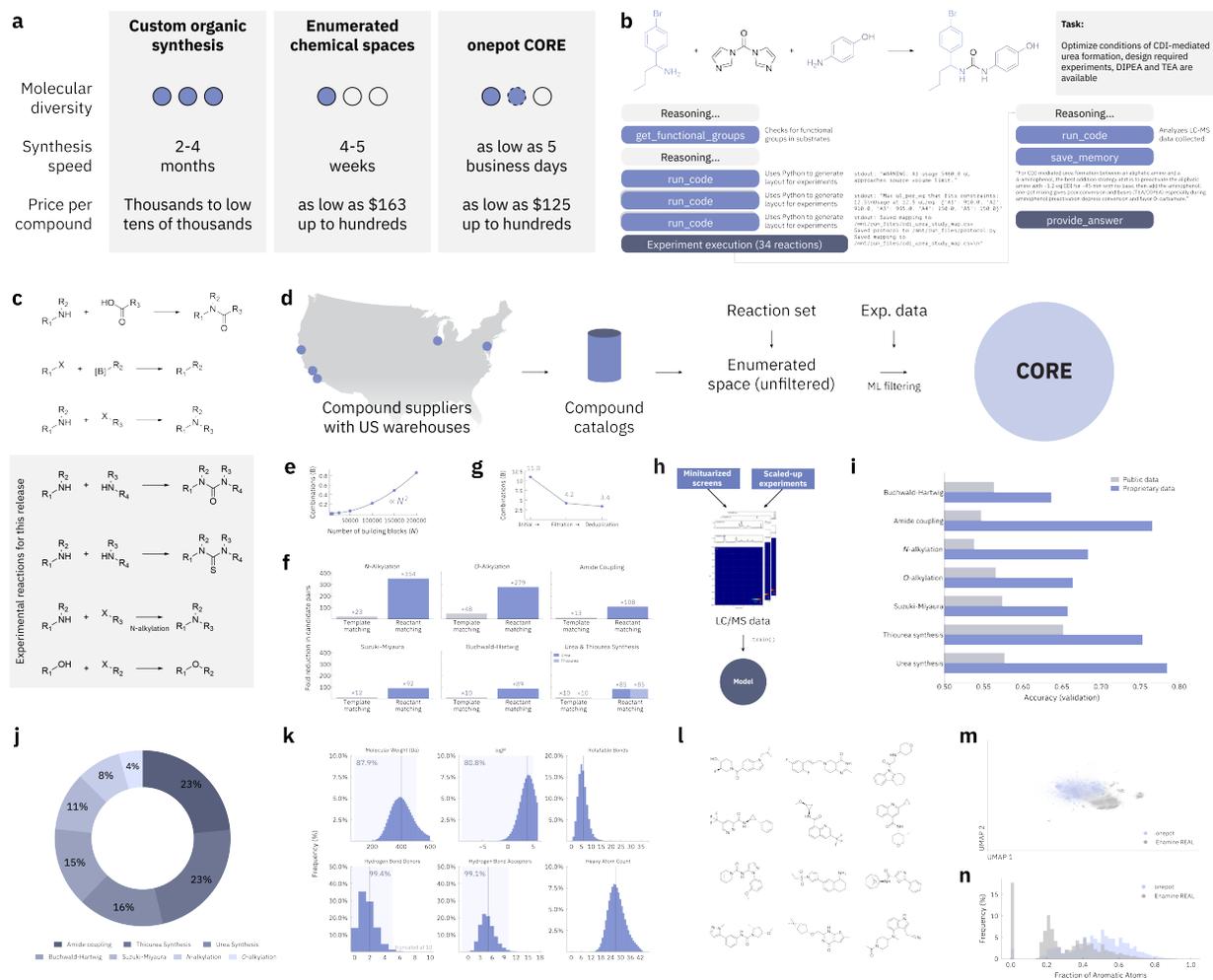

**Figure 1.** Chemical space construction process. **a,** Comparison between different small molecule synthesis options. **b,** An example protocol execution trace for urea synthesis. **c,** Reactions used to construct the space in this release. **d,** Overview of space construction strategy. **e,** Space size scaling with increasing number of building blocks for Buchwald-Hartwig coupling. **f,** Performance optimizations for enumeration across selected reaction classes. **g,** Chemical space size progression. **h,** ML model development workflow. **i,** Comparison of ML models performance with respect to data sources used for training. **j,** Distribution of reaction classes in CORE. **k,** Distribution of key molecular properties; shaded regions in subplots correspond to Lipinski's rule of five compliant regions. **l,** Sample of CORE molecules with high QED scores. **m,** Visualization of CORE and Enamine REAL spaces. **n,** Distribution of fractions of aromatic across spaces.

# Constructing the space

To construct the space, we follow a relatively straightforward strategy — identifying reactions of interest, sourcing building blocks, enumerating all the compounds in the space, and performing ML-based feasibility assessment.

### *Reaction set*

In this release we are providing seven reactions that are widely used in medicinal chemistry (**Figure 1c**). Three of them have been already tested with our commercial partners: amide coupling, Suzuki-Miyaura coupling, and Buchwald-Hartwig amination. Four other reactions are being beta-tested in this release: urea synthesis, thiourea synthesis, amine alkylation, and O-alkylation.

Protocol development and condition optimization for these reactions were performed by our LLM agent *Phil*, which had access to an extensive set of tools to design, execute, and analyze the results of the experiments. For example, in **Figure 1b** we show one of the execution traces for development of one of the protocols for CDI-mediated urea synthesis: *Phil* performs initial analysis, designs experiment protocols for execution, executes them on physical hardware, and then analyzes the data to develop a refined protocol as well as store knowledge. We aim to further extend our reaction portfolio with three reactions in testing and more in preparation.

The developed protocols are then tested with a broad variety of different substrates to learn their applicability scope. Scaled-up syntheses are also fed in this system to further improve the platform.

When selecting the reactions for the chemical space creation, it is important that the resulting molecules are diverse enough, and the space is large enough for downstream applications such as virtual screening. The reactions that were selected for this version of the space satisfy both criteria — they operate on a broad scope of available starting materials with each reaction class providing significant number of molecules to the final chemical space composition resulting in high diversity of CORE.

### *Building blocks*

Building a large stock of building blocks internally would be a challenging task — having diverse stock would require tens of millions of dollars investment, with no guarantee that any significant fraction of them will actually be used in customer synthesis. Instead, we have identified a set of trusted suppliers with large stock of compounds in the US, built direct relationships with them, and validated their performance. As a result, we construct a meta-catalog with full current availability, prices, and historical performance of that supplier.

All building blocks across different suppliers are aggregated and filtered to remove the highly reactive molecules, isotopically labeled building blocks, or molecules that have very high molecular weight. Supplier risk and prices of the building blocks are later incorporated into the catalog during the enumeration process.

### *Enumeration*

Once building blocks and the reaction set are identified, onepot performs full enumeration of all possible products in these reactions. The task is computationally challenging due to quadratic scaling of possible reactant combinations with respect to the number of building blocks used for enumeration (**Figure 1e**), requiring a significant number of CPU hours to complete.

Enumeration begins with taking the set of SMILES of all the accessible building blocks and combines them with the SMIRKS templates for the reactions that are used for the chemical space creation. Building blocks and templates are used to construct the fully enumerated version of the catalog. Due to the scale of enumeration, it needs to be done in a distributed way. There are several optimizations that are done at this stage that reduce the amount of computation required: for every reaction we group the building blocks by matching them with SMARTS templates of reactants that constitute given reaction template. This way we only enumerate the valid pairs of reactants, which gives significant boost in performance compared to (i) considering all possible reactant pairs and (ii) considering all possible pairs of the reactants that matched at least one reactant SMARTS pattern (**Figure 1f**).

This first stage provides the superset of all the reactions that will be included in the final version of the chemical space, but it is only the first step in the process — the space is still far from being useful — it still contains duplicate products. Additionally, SMIRKS templates are not precise enough to capture the full scope for a particular reaction class.

*Feasibility assessment*

Even if the substrates match the template, it does not mean that the reaction will actually proceed. Instead, researchers usually implement a number of manual filters to account for incompatible functional groups and try to make the reaction template as specific as possible. However, this approach can lead to a significant number of both false positives and false negatives. On false positives side, some of the effects might be hard to predict; furthermore, the space of possible functional groups and atomic arrangements is quite large, and it is hard to manually define all the possible failure modes. False negatives are also a problem; sometimes filters are just too strict and constraining synthesizable space. Creation of the manual filters also makes the addition of reactions considerably more complicated — each new reaction requires a new set of hand-crafted filters constructed from existing knowledge or analysis of experimental data, which is typically done by human chemists and, therefore, naturally takes quite some time.

ML-based feasibility assessment offers a solution here — instead of manually defining filters, we collect large datasets of running each specific protocol and then evaluate its applicability scope using these models. The ML-based filtration provides a very significant reduction in terms of the SMILES that are kept in the final version of CORE (**Figure 1g**).

To evaluate the models before filtration, we curate 3 different sets of evaluation sets. The "simplest" set contains the reactions for which both reactants were present in the training for a particular reaction class; the "medium"-difficulty evaluation set contains the reactions for which exactly one of the reactants was not present in the training dataset for a selected reaction class. Finally, the most difficult evaluation set contains the reactions for which both building blocks were not present in the training dataset at all. The last set evaluates the ability of a model to generalize novel substrates.

Besides letting us compare the relative strengths of different models, the evaluation set also lets us recalibrate our model's confidence in success with respect to actual success probability. As such, we can then appropriately set confidence thresholds for filtration based on the true success probability we wish to see in the lab. This calibration also allows us to categorize all feasible candidate reactions into three categories: low, medium, or high chemical risk. These categories are created via bucketing based on our confidence levels.

We use multiple tiers of data to get the highest performance: pre-training scale datasets, LC/MS miniaturized screenings, and actual full-scale syntheses (**Figure 1h**). In this order, the number of datapoints decreases, but data becomes less noisy and more relevant to actual commercial syntheses. Internally collected data provides a notable increase in the model's performance across all reaction classes (**Figure 1i**) on the joined evaluation set. For example, amide coupling saw accuracy increase from 55% to 77% including challenging substrates. It is important to note that the accuracies are calculated on balanced sets of reactions, and actual success rate depends on overall protocol performance as well.

### *Final space assembly*

Final step of the chemical space creation includes aggregation of all the files from the previous stage into a single file, deduplication, price calculation and assignment of coarse-grained feasibility score to every molecule in the final version of the chemical space. The size of the final version of the chemical space is considerably smaller than the initial version right after enumeration stage (**Figure 1g**).

The chemical space includes SMILES of the product, identifier of the molecule, supplier risk score, and chemistry risk score, therefore, giving as much context to the users when working with CORE.

### *Space composition*

Most common reaction class in the CORE space is amide coupling demonstrating broad scope of the reaction and high experimental success rate (**Figure 1j**). We noticed that even though Buchwald-Hartwig has a larger scope than amide coupling, its contribution is considerably smaller due to the higher reaction sensitivity that is captured and filtered by our models.

The resulting molecules have drug-like properties with approximately 73% of compounds obeying Lipinski's rule of five (**Figure 1k**). The sample of the molecules with high QED values are demonstrated in **Figure 1l**.

Finally, we compare onepot CORE space with Enamine REAL (**Figure 1m**) by plotting the UMAP projections of Morgan fingerprints for both spaces (by taking a sample of 400,000 molecules from each of the spaces). The results demonstrate that the spaces differ considerably from each other — we found that REAL contains a much larger fraction of molecules, containing primarily non-aromatic carbon atoms, than CORE (**Figure 1n**) leading to the difference on the building blocks level.

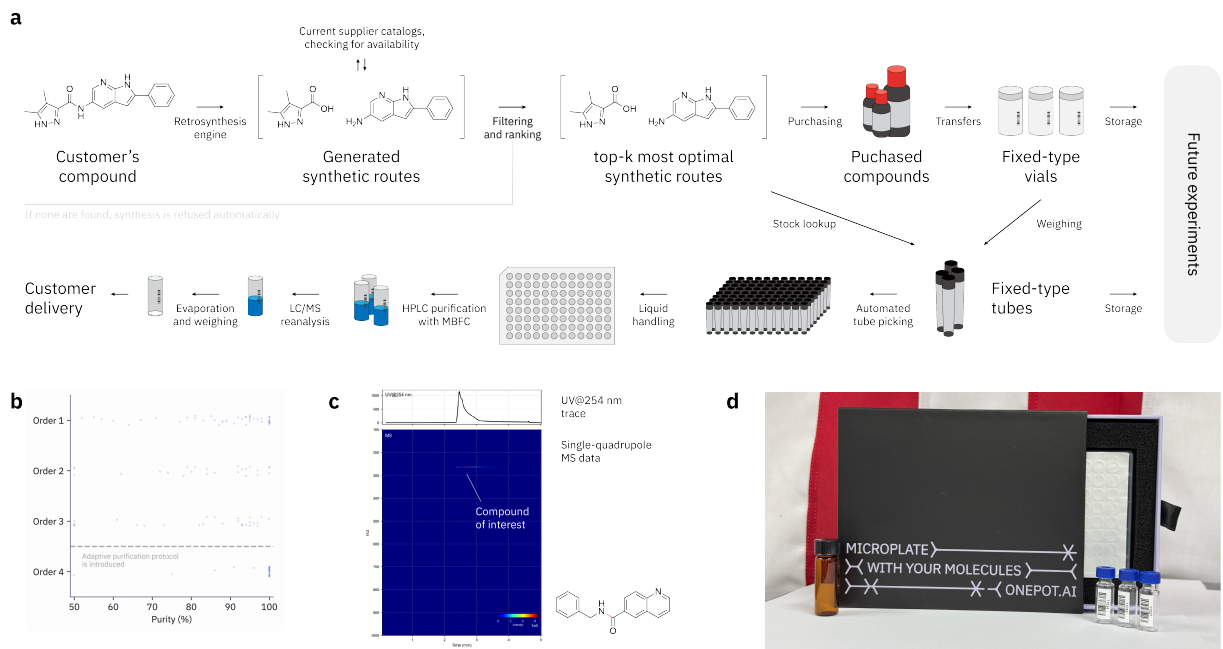

**Figure 2.** Synthesis strategy. **a,** Overview of the synthesis process. **b,** Purity statistics. **c,** Example analytical data provided with the compounds. **d**, Final shipments.

## Executing the syntheses

After the space was constructed, we tested our ability to execute these experiments in our automated synthesis platform. In this section we give a high-level overview of the synthesis process for a compound from space (**Figure 2a**).

When a compound is selected, we pick available routes for synthesis of this compound. In some cases, same reactions can be used with multiple substrates (for example, Cl-, Br-, I-substituted substrates in Suzuki coupling), or different reactions. Then we look up missing building blocks for each of these reactions, using supplier catalogs. If none of them are available, we cannot proceed with the synthesis, otherwise we pick at least one (often more to collect data and derisk syntheses) route based on model predictions and our internal heuristics and order corresponding building blocks.

Once building blocks are received, they are unpacked, registered in the system, and transferred to a fixed-format vials for long-term storage, as well as sampled to prepare a stock solution in a tube. Both vials and tubes are designed to be compatible with automated storage systems.

Next, the stock solution tubes are picked and transferred into a liquid handler, corresponding protocols are combined into a single sequence for transfers on these liquid handlers. Depending on the reaction, some of them will be set up using a liquid handler in a glovebox, some reactions will require reactive on-deck reagents, that will be prepared separately. The reaction mixtures are then heated if needed, and basic workup is performed.

After that, resulting crude materials are sent for purification on semi-preparative scale HPLC system with mass-based fraction collection. We use an internal tool for automated protocol generation for reaction mixtures; the tool helped us to achieve a considerable increase in purification success rates even for challenging substrates (**Figure 2b**). The fractions are collected in pre-tared barcoded vials.

Finally, vial solutions are reanalyzed again using HPLC/MS system with single quadrupole mass analyzer. Purity is confirmed based on UV signal at 254 nm (**Figure 2c**). Solvent is removed using either a $N_2$ blow-down evaporator or a centrifugal evaporator. Final compound is delivered in vials as dry solid or in plates as 10 mM DMSO solution (**Figure 2d**).

This flow enables us to synthesize compounds very quickly giving turnaround of just a few days. The capabilities of the system will be further improved, with photochemical and electrochemical setups on a roadmap.

## Validation

### *Timelines*

onepot CORE has a unique advantage — we operate in the United States, so US-based users do not have to wait for customs clearance, which in practice adds at least a week to timelines, and we also do not have to wait for our reagents to clear customs either. When combined with automated synthesis platform, timelines become particularly impressive — as fast as five business days.

In the very first order, 20/50 compounds were synthesized in three business days, 40/50 in 4 more days, and remaining amount in 5 more days. Synthesis itself took 1-2 days; the remaining time was spent waiting for building block suppliers. These delays led to exclusion of a number of suppliers from the platform, to increase shipment reliability. One of the most recent orders with 52 compounds synthesized was done in 9 business days in late December. This order required extensive condition optimization and use of air-free experimentation techniques. The fastest order end-to-end contained 30 compounds, synthesized and shipped in 5 business days.

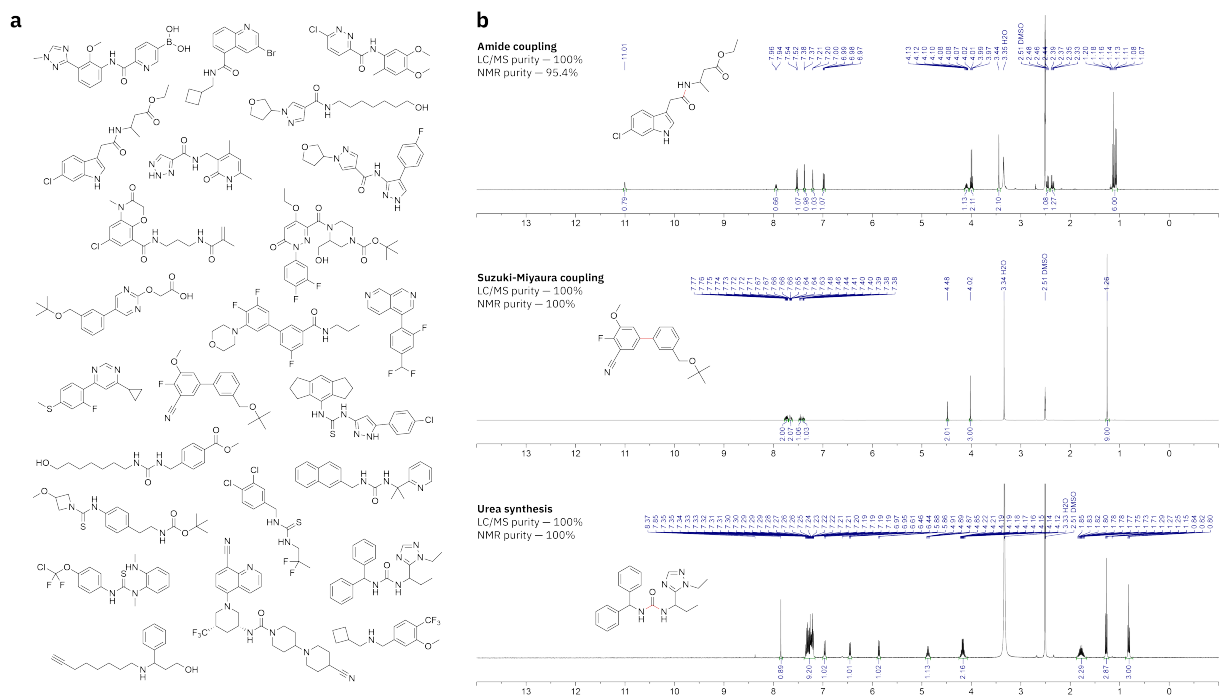

**Figure 3.** Compound identity confirmation results. **a,** Example structures synthesized for compound identity confirmation. **b**, Example experimental NMR spectra of compounds from the space.

## *Compound identity and purity*

When relying on pretty much unit mass from single quadrupole mass analyzer to confirm compound identity, one of the most common concerns is presence of isomers or isobaric compounds. We prioritize chemistries expected to yield a single major product; when multiple products are detected, we report them explicitly and (where feasible) isolate them as separate fractions. Furthermore, as a part of validation process, we synthesized 24 compounds (**Figure 3a**) on our platform and checked their identity using [1]H NMR. The set of compounds is very diverse and contains various functionalities. The spectra were collected in DMSO-d6, using a 400 MHz instrument.

For all these compounds, [1]H NMR spectrum was consistent with the expected structure supported by LC/MS and reaction constraints. **Figure 3b** contains three examples for amide coupling, Suzuki-Miyaura coupling, and urea synthesis.

In the future, we aim to use tandem mass spectrometry data to provide structure confirmation for all the synthesized compounds. Even though the method often provides less conclusive results than NMR, we aim to correct that using ML models and priors about reaction mixture composition.

Despite emergence of such fields as direct-to-biology experimentation, for many biological assays compound purity is still a critical factor. On our platform, every compound is accompanied by an LC/MS report, confirming compound identity. However, some species might not come off column at all, some of them will not ionize, and some will not absorb UV signal. In the future, we aim to add other detectors, such as CAD or ELSD.

We used data from the identity confirmation study to compare analytical results and experimental spectra. In the examples, shown in **Figure 3b** NMR did pick up a signal of likely a starting material for the amide coupling product. However, this did not decrease expected purity much — while LC/MS has shown only a single peak, so assumed purity is 100%, proton-count weighted ratio of integrals, gives 95.4% expected purity. Some of the compounds have a singlet at 8–8.5 ppm, corresponding to formic acid or formate counterion. Finally, depending on evaporation method fresh compound can contain residual amounts of water as seen in the NMR spectra (even though DMSO solution can absorb it as well). We have implemented protocols to keep compounds dry, which involves using longer evaporation methods and keeping them in desiccator before weighing and shipment. No other impurities were observed for these compounds.

*Biological assay validation*

Finally, even though the compound identity and purity are confirmed, can these compounds be used in biological assays? To validate this, we have prepared a series of DPP4 inhibitors and performed corresponding assays.

DPP4 breaks down a number of hormones, including GLP-1. Its inhibitors were explored as potential drugs for type 2 diabetes. To showcase the platform, we used sitagliptin, vildagliptin, and saxagliptin as compounds around which we explored SAR (**Figure 4a**). Initially, we input them at onepot.ai website to find similar analogs, these analogs were then decomposed, and full matrix of pairwise substituents was constructed (**Figure 4b**). As a final step we needed to perform deprotection — this combination of amide coupling and deprotection is currently being extensively tested and is available if requested. These compounds were then subjected to commercially available DPP4 fluorescence assay.

The results of this study are presented in **Figure 4c**. IC$_{50}$ for commercially available standard and resynthesized compound are very close. Obviously, drug discovery process involves consideration of a number of other factors, including, but not limited to physicochemical properties, ADME properties, and toxicity. However, this study demonstrates how onepot CORE can be used for effective SAR exploration.

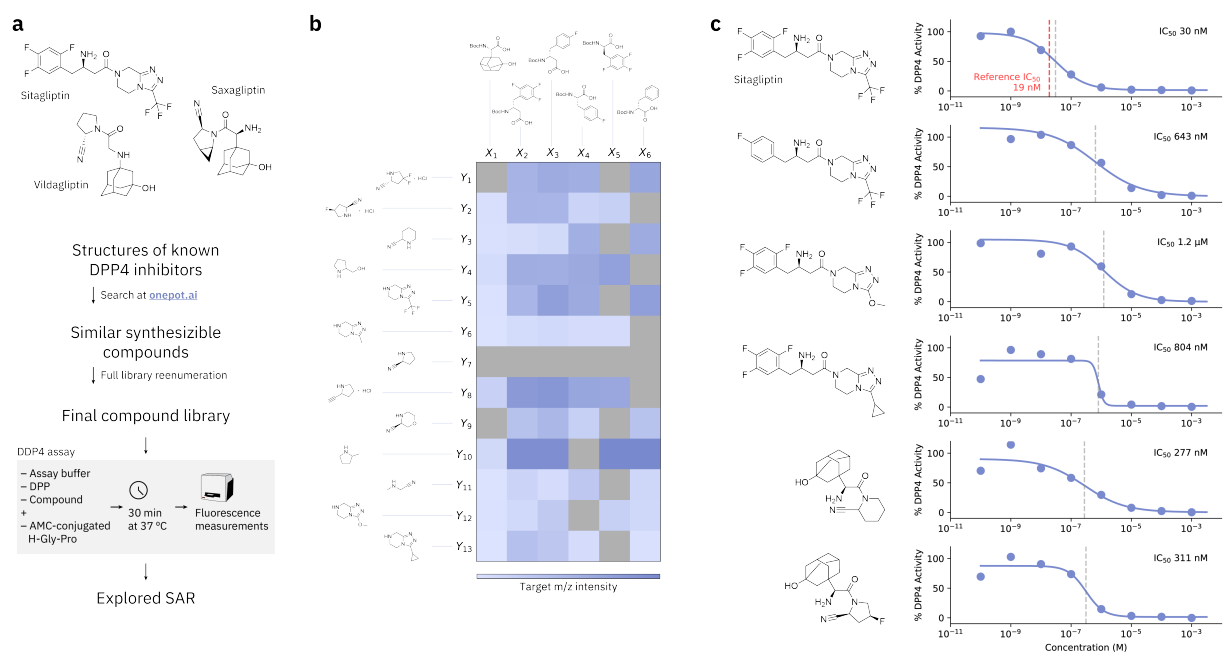

**Figure 4.** Biological assay validation. **a,** Overview of the workflow. **b,** Experimental synthesis results. **c,** Assay results, obtained with compounds synthesized using the platform.

## Conclusions

In this work, we described onepot CORE — an enumerated chemical space containing 3.4B compounds, enabled by reaction protocols, developed by *Phil*, our AI chemist. The chemical space is enabled by complete and very tight integration of the entire process — compound availability estimation, enumeration, reaction outcome prediction, experimental execution from weighing to purification, and final QC and analysis. Furthermore, we performed extensive testing by verifying compound identity and purity, their applicability in biological assays, and also commercially through observed purities and execution timelines.

It is hard to overestimate impact of truly automated small molecule synthesis. The advantages are two-fold. First, such platforms, when done right, considerably increase iteration speed. For instance, our partners can perform five iterations of compound selection in the same timeframe as they could do with other suppliers. Even under the same budget, faster iterations are likely to bring them to better molecules. Second, using ML-based filters, one can take higher chemical risk when executing experiments. Instead of blindly filtering out entire sets of functional groups, one can build more detailed and fine-grained understanding of compound reactivity, functional group tolerance, and reagent/catalyst performance. This opens a door to even broader chemical space, that could contain many of the drugs of the future.

We will continue improving the platform, bringing in new reactions, one-pot reaction sequences, and multistep chemistry to further make chemical synthesis more accessible. We believe that organic chemistry is still very far from being "solved", the challenges involve precise control of reactivity (e.g., being able to C-H-activate any bond in a molecule would be a Nobel-level achievement), better understanding of reaction mechanisms, and most importantly for us, extending scope of reactions that go cleanly under mild conditions — moving beyond amide coupling, CuAAC, and SuFEx chemistries.

## Conflicts of Interest

All authors are employees of Onepot AI, Inc. and may hold equity or equity-linked compensation in the company. Onepot AI, Inc. develops and commercializes technologies related to the subject matter of this manuscript. Any intellectual property arising from this work may be owned by, or assigned to, Onepot AI, Inc.